\documentclass[12pt]{article}

\usepackage{scicite}

\usepackage{times}

\usepackage{graphicx}
\usepackage{slashed}
\usepackage{amssymb}
\usepackage{extarrows}
\usepackage{amsmath}
\usepackage{braket}

\usepackage{xcolor}  

\newcommand{\Yb}{\ensuremath{^{171}\mathrm{Yb}^+~}}

\newenvironment{sciabstract}{
\begin{quote} \bf}
{\end{quote}}

\newcounter{lastnote}

\title{Time Reversal and Charge Conjugation\\ in an Embedding Quantum Simulator}

\author
{Xiang Zhang$^{1}$, Yangchao Shen$^{1}$, Junhua Zhang$^{1}$, \\Jorge Casanova$^{2,3}$, Lucas Lamata$^2$, Enrique Solano$^{2,4}$, \\Man-Hong Yung$^{1}$, Jing-Ning Zhang$^{1}$, Kihwan Kim$^{1}$ \\
\\
\normalsize{$^{1}$Center for Quantum Information, Institute for the Interdisciplinary Information Sciences,}\\
\normalsize{ Tsinghua University, Beijing, 100084, P. R. China}\\
\normalsize{$^{2}$Department of Physical Chemistry, University of the Basque Country UPV/EHU,}\\
\normalsize{ Apartado 644, E-48080, Bilbao, Spain}\\
\normalsize{$^{3}$Institut f\"ur Theoretische Physik, Albert-Einstein-Allee 11, Universit\"at Ulm, D-89069 Ulm, Germany}\\
\normalsize{$^{4}$IKERBASQUE, Basque Foundation for Science, Maria Diaz de Haro 3, 48013 Bilbao, Spain} \\
}

\date{}

\begin{document}

\baselineskip24pt

\maketitle

\begin{sciabstract}
The understanding of symmetry operations has brought enormous advancements in physics, ranging from elementary particle to condensed matter systems. In quantum mechanics, symmetry operations are described by either unitary or antiunitary operators, where the latter are unphysical transformations  that cannot be realized in physical systems. So far, quantum simulators of unitary and dissipative processes, the only allowed physical dynamics, have been realized in key experiments. Here, we present an embedding quantum simulator able to encode unphysical operations in a multilevel single trapped ion. In this sense, we experimentally observe phenomena associated with the nonunitary Majorana dynamics and implement antiunitary symmetry operations, i.e., time reversal and  charge conjugation, at arbitrary evolution times. These experiments enhance the toolbox of quantum simulations towards applications involving unphysical operations.
\end{sciabstract}

A quantum simulator is a promising solution to solve a type of problems that are intractable with classical means \cite{Feynman82, Lloyd96}. The quantum simulator \cite{CiracZoller12} is expected to revolutionize various disciplines that require demanding computations including condensed matter physics \cite{Porras04,Kim10}, quantum chemistry \cite{Aspuru-Guzik05,Lanyon10,Yung14},  relativistic quantum mechanics \cite{Lamata07,Gerritsma10}, quantum field theory \cite{Casanova11b,Jordan12} and lattice gauge theories \cite{Reznik13,Banerjee13}. In recent years, a variety of physical platforms such as neutral atoms \cite{Bloch12}, ions \cite{Blatt12}, photons \cite{Aspuru-Guzik12}, and superconducting circuits \cite{Koch12} have been fruitfully developed to simulate many of the physical processes occurring in nature. Up to now, the main endeavors of the quantum simulator have been focused on efficiently simulating physical processes. Indeed, diverse physically meaningful transformations, as symmetry operations, are impossible to happen in nature or to be observed in the laboratory. Nevertheless, we are used to compute these unphysical transformations with classical resources for the sake of useful scientific calculations.

The study of the above mentioned symmetries has profoundly shaped our comprehension of physical laws. Wigner proved that any symmetry operation acts as a unitary or antiunitary transformation in the Hilbert space, which is known as Wigner's theorem \cite{GroupTheory}. Charge conjugation and time reversal are paradigmatic examples of antiunitary operations \cite{SakuraiAdvanced}. The charge-conjugation operation converts a particle into its antiparticle, which changes the sign of the charge. The time-reversal operation inverts the motion \cite{SakuraiModern}, which effectively reverses the direction of the time axis. The charge conjugation together with the parity symmetry is not conserved in the weak interaction \cite{TDLeeCPViolation,KleinknechtCPViolation}, just as the time-reversal symmetry. The discovery of the violation of these symmetries has been a decisive breakthrough of the standard model.

The Majorana equation~\cite{Majorana},
\begin{equation}
\label{MajoranaEquation}
i \hbar\slashed{\partial}\psi = m c \psi_c ,
\end{equation}
where $\psi$ and its charge conjugate $\psi_c$ are both present, describes a fundamental non-Hamiltonian system. Similar to the Dirac equation, the Majorana equation satisfies the Lorentz covariance required by relativistic quantum mechanics. Majorana envisioned that the real version of Eq.~(\ref{MajoranaEquation}), in which $\psi=\psi_c$, would be the fundamental equation describing neutrinos~\cite{Majorana}. Although it is still an open question whether neutrinos are Dirac or Majorana particles, the complex Majorana dynamics of Eq.~(\ref{MajoranaEquation}) has its own theoretical importance in exploring physics beyond the standard model. Due to the presence of the complex conjugation operation, the Majorana dynamics possesses unique features, such as nonconservation of charge and momentum, broken orthogonality, and nontrivial effect of the state global phase. Recently, a quantum simulation of the Majorana dynamics was performed in a photonic quantum platform, by decomposing its evolution in two Dirac equations~\cite{MajoranonTheory,MajoranaOptical}.

\begin{figure}[htbp]
  \includegraphics[width=\textwidth]{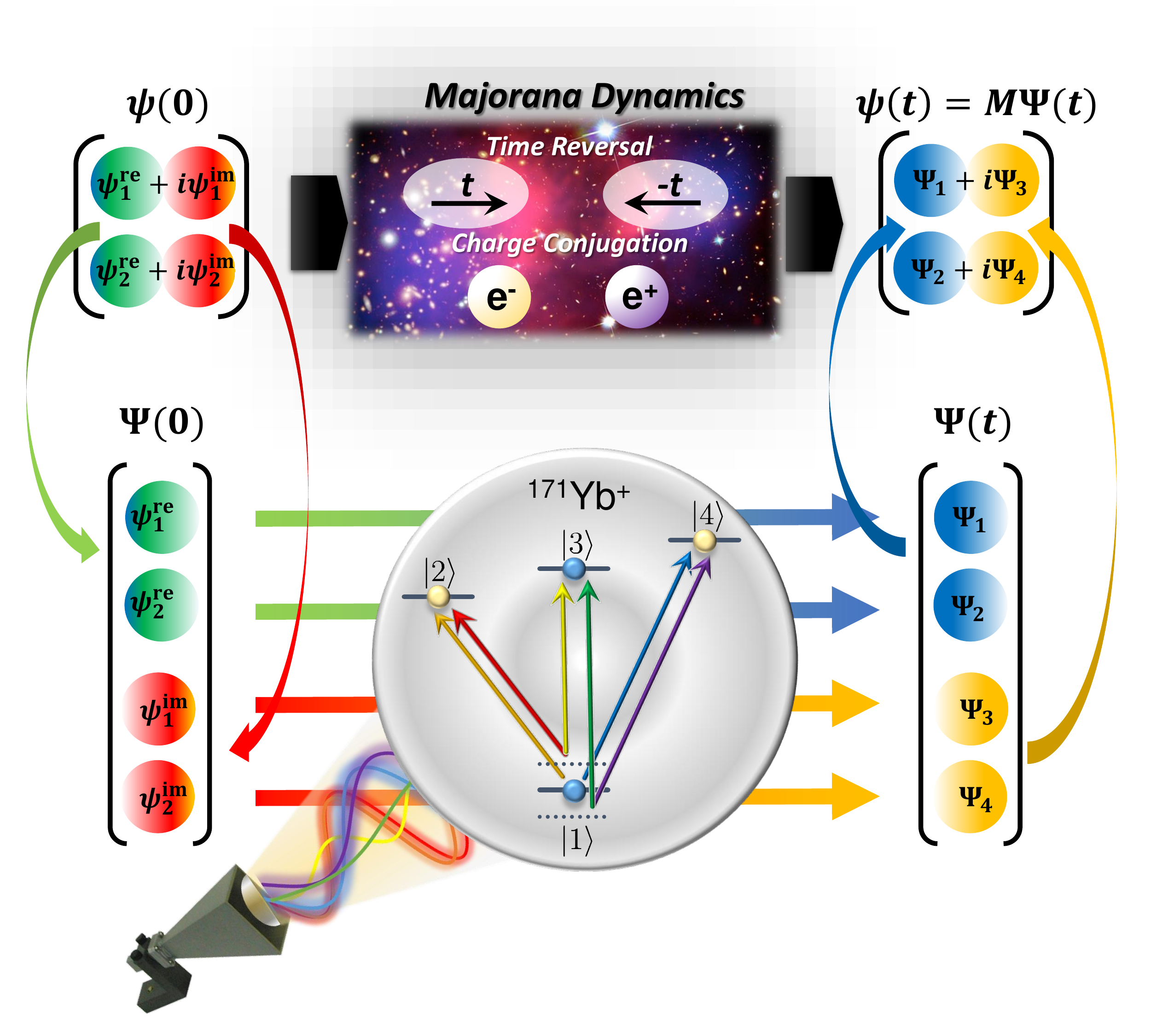}\\
  \caption{\textbf{Schematic of the embedding quantum simulator.} The upper and lower parts represent the original and enlarged spaces, respectively. Unphysical processes, which are forbidden by the laws of quantum mechanics, are mapped to unitary operations in the enlarged space. The embedding quantum simulator is built in a single $^{171}{\rm Yb}^+$ ion trapped in a linear Paul trap, where the enlarged space is encoded in the ground-state manifold of the ion. The unitary operations are implemented by applying microwaves with six frequencies from a microwave horn.\label{fig:Fig1}}
\end{figure}

In 2011, it was discovered that unphysical operations can be implemented in an embedding quantum simulator (EQS) by enlarging the associated Hilbert space~\cite{Casanova11a}. The proposed scheme enables us  to explore paradigmatic unphysical operations, such as time-reversal and charge-conjugation symmetries. This scheme can be also applied to measure entanglement monotones without full state tomography \cite{Candia13} or to perform noncausal kinematic transformations \cite{Rodriguez13}. An EQS mapping ${\mathcal M}:{\mathbb C}^2\rightarrow{\mathbb R}^4$ may be defined as
\begin{eqnarray}
\psi=\left(\begin{array}{c} \psi_1^{\rm re}+i \psi_1^{\rm im}  \\ \psi_2^{\rm re}+i \psi_2^{\rm im} \end{array}\right)\xlongrightarrow{\mathcal M}\Psi=\left(\begin{array}{c} \psi_1^{\rm re} \\ \psi_2^{\rm re} \\ \psi_1^{\rm im} \\ \psi_2^{\rm im} \end{array}\right),\label{eq:Mapping}
\end{eqnarray}
where ${\mathbb C}^2$ and ${\mathbb R}^4$ refer to the original and enlarged Hilbert spaces, respectively. Through the mapping ${\mathcal M}$, the complex conjugation $\hat K:\psi\rightarrow\psi^*$, the time reversal $\hat T:t\rightarrow\left(-t\right)$, and the charge conjugation $\hat C:\psi\rightarrow\psi_c$ take the form of unitary two-qubit gate operations in the enlarged Hilbert space: $\hat{\mathcal K}=\hat\sigma_z\otimes\hat{\mathbb I}$, $\hat{\mathcal T}=i\hat\sigma_z\otimes\hat\sigma_y$, and $\hat{\mathcal C}=-\hat\sigma_z\otimes\hat\sigma_x$, respectively. As shown in Fig.~\ref{fig:Fig1}, one can recover the quantum state in the original Hilbert space, $\psi$, via the use of the matrix $M$ acting on $\Psi$.

The Majorana equation in $1+1$ dimensions, which is given by
\begin{eqnarray}
i\hbar\partial_t\left|\psi\right\rangle=\hat H_M\left|\psi\right\rangle=\left(c\hat\sigma_x\hat p_x-imc^2\hat\sigma_y\hat K\right)\left|\psi\right\rangle,\label{eq:MajoranaEquationOriginal}
\end{eqnarray}
is mapped, through the EQS transformation of Eq.~(\ref{eq:Mapping}), to an effective Hamiltonian system
\begin{eqnarray}
i\hbar\partial_t\left|\Psi\right\rangle=\hat{\mathcal H}\left|\Psi\right\rangle=\left[c\hat p_x\left(\hat{\mathbb I}\otimes\hat\sigma_x\right)-mc^2\left(\hat\sigma_x\otimes\hat\sigma_y\right)\right]\left|\Psi\right\rangle,\label{eq:MajoranaEquationEnlarged}
\end{eqnarray}
which can be directly implemented in a physical system.

The ion-trap system, a leading platform for quantum simulation \cite{Blatt12}, is used to implement the unphysical dynamics and operations. Our system consists of a single \Yb ion confined in a linear Paul trap \cite{Contextuality13}. The ground-state manifold $^2S_{1/2}$ contains four internal states denoted by $\left|F=0,m_F=0\right\rangle\equiv\left|1\right\rangle$ and $\left|F=1,m_F=-1,0,1\right\rangle\equiv\left|m_F+3\right\rangle$, which are separated by the hyperfine splitting $\omega_{\rm HF}=\left(2\pi\right)12.642{\rm GHz}$. A uniform static magnetic field $B= 9.694 {\rm G}$ is applied to define the quantization axis and causes Zeeman splitting $\omega_Z=\left(2\pi\right)13.5855 {\rm MHz}$.

After transforming Eq. (\ref{eq:MajoranaEquationEnlarged}) to the momentum space, we obtain a simpler Hamiltonian $\hat{\mathcal H}_p=pc\left(\ket{1}\bra{2}+\ket{3}\bra{4}\right)+i mc^2 \left(\ket{1}\bra{4}-\ket{2}\bra{3}\right) + {\rm H.c.}$, where the momentum operator $\hat{p_x}$ is substituted by its eigenvalue $p$ (see Supplementary Materials). For simplicity, we use a set of dimensionless units, i.e. $mc^2$ for the energy, $mc$ for the momentum, and $\frac{\hbar}{mc^2}$ for the time.

The experimental procedure is as follows. First, we map an initial Majorana spinor $\psi(x,t=0)$ to a real bispinor $\Psi(x,t=0)$ in the enlarged space. The momentum representation of the bispinor $\widetilde\Psi\left(p,t=0\right)\equiv\frac{1}{\sqrt{2\pi}}\int\Psi\left(x,0\right)e^{-ipx/\hbar}dx$ evolves according to the enlarged space Hamiltonian $\hat{\mathcal H}_p$. After encoding $\widetilde\Psi\left(p,0\right)$ into the ground-state manifold of the trapped $^{171}{\rm Yb}^+$ ion, we apply microwaves with six frequencies to implement $\hat H_p$. In particular, we use resonant microwaves to couple $\left|1\right\rangle\leftrightarrow\left|2\right\rangle$ and $\left|1\right\rangle\leftrightarrow\left|4\right\rangle$, and the stimulated two-photon Raman processes with opposite detunings to couple $\left|2\right\rangle\leftrightarrow\left|3\right\rangle$ and $\left|3\right\rangle\leftrightarrow\left|4\right\rangle$ as shown in Fig. \ref{fig:Fig1}. After evolving for a certain time $t$, we perform quantum state tomography to obtain the enlarged space density matrix $\hat\varrho\left(p,t\right)$, which can be mapped to the original space density matrix $\hat\rho\left(p,t\right)$. The average value of a diagonal operator $A_d$ in the momentum space can be directly obtained via integration over the momentum, $\left\langle\hat A_d\right\rangle=\int{\rm Tr}\left[\hat A_d\hat\rho\left(p,t\right)\right]dp$.
To obtain the average value of an off-diagonal operator in the momentum space, for example the average position of the Majorana particle, we change Eq. (\ref{eq:MajoranaEquationEnlarged}) into a pair of decoupled two-dimensional equations by diagonalizing the first qubit. By coherently evolving a couple of two-dimensional equations with different momenta, we obtain the phase information between different momentum components. We repeat each measurement 1000 times to get the expectation value. The statistical errors, which are mainly due to quantum projection, are estimated by the standard deviation of mean value.

\begin{figure}[htbp]
  \includegraphics[width=\textwidth]{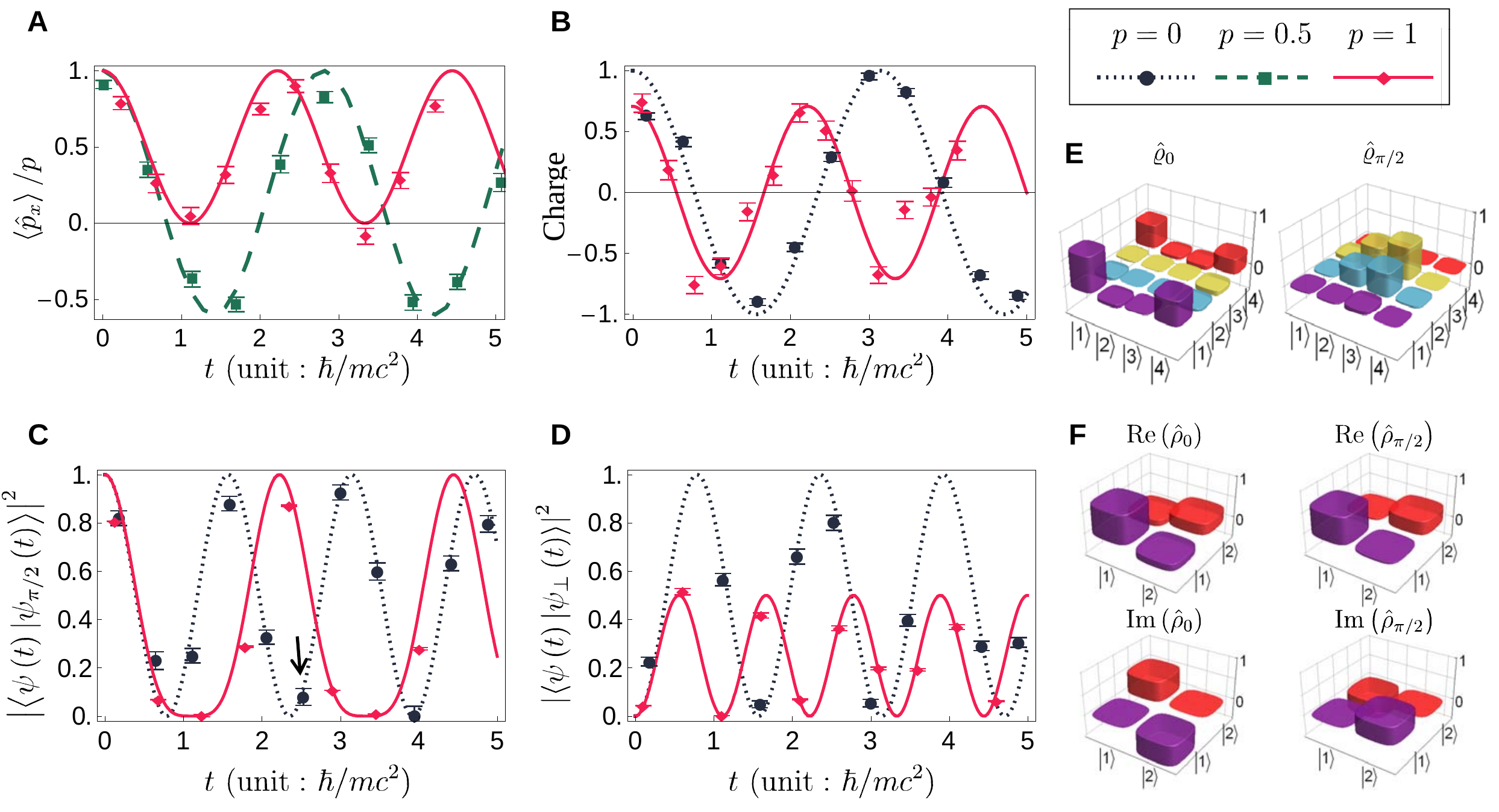}\\
  \caption[Text excluding the matrix]{\textbf{Majorana dynamics}. ({\bf A}) Momentum space {\it Zitterbewegung}. ({\bf B}) Violation of charge conservation. The average values of the physical observables in ({\bf A}) and ({\bf B}) are measurement results of the Majorana spinor $\left|\psi\left(t\right)\right\rangle$ evolving from the initial state $\left(\begin{smallmatrix} 1 \\ 0 \end{smallmatrix}\right)\otimes\left|p\right\rangle$. ({\bf C}) Nonconserved fidelity caused by an initial global phase. The Majorana spinors $\left|\psi_\theta\left(t\right)\right\rangle$ evolve from the initial states $e^{i\theta}\left(\begin{smallmatrix} 1 \\ 0 \end{smallmatrix}\right)\otimes\left|p\right\rangle$ with $\theta=\pi/2$. ({\bf D}) Nonconserved orthogonality for initially perpendicular Majorana spinors. The Majorana spinors $\left|\psi_\perp\left(t\right)\right\rangle$ evolve from initial states $\left(\begin{smallmatrix} 0 \\ 1 \end{smallmatrix}\right)\otimes\left|-p\right\rangle$. We choose the momenta of the initial plane-wave states as $p=0$ (black dotted), $0.5$ (green dashed), and $1$ (red solid), and set the Majorana mass $m=1$. Curves are from theoretical simulation, and dots are from experimental data. ({\bf E}) Density matrices in the enlarged space obtained by quantum state tomography, related to the data point marked by the black arrow in ({\bf C}). ({\bf F}) Reconstructed density matrices in the original space. Error bars, $1\sigma$.\label{fig:Fig2}}
\end{figure}

Fig.~\ref{fig:Fig2} shows our experimental results of the Majorana dynamics, where the initial spinors are chosen to be plane-wave states with $\psi_2=0$, i.e. $\left|\psi\left(0\right)\right\rangle=\left(\begin{smallmatrix} 1 \\ 0 \end{smallmatrix}\right)\otimes\left|p\right\rangle$. Fig.~\ref{fig:Fig2} ({\bf A}) shows the momentum space {\it Zitterbewegung} for a Majorana particle. Due to the existence of the complex conjugate operator in the Majorana equation, the momentum, which is conserved for free Dirac particles, is no longer a conserved quantity in the Majorana dynamics. Because the violation of momentum conservation is originated by the Majorana mass term, the amplitude of the oscillation is inversely proportional to the magnitude of the momentum of the initial state. Meanwhile, the frequency of the oscillation is determined by the relativistic dispersion relation $\sqrt{p^2+m^2}$, so the initial plane wave with larger momentum will oscillate faster. As shown in Fig.~\ref{fig:Fig2} ({\bf B}), the Majorana dynamics also violates charge conservation, which may lead to physics beyond the standard model \cite{MassiveNeutrinos}. In the rest frame, the charge operator measures the difference between the populations of the internal states, which is equivalent to the $\hat\sigma_z$ operator \cite{LLamata12}. For the non-zero momentum case, the particle and antiparticle basis is obtained by diagonalizing the corresponding Dirac equation with the same momentum, and the charge of a Majorana spinor is defined as the difference between the populations of the particle and antiparticle components (see Supplementary Materials). For the same reason, the amplitude and frequency of the charge oscillation exhibits similar momentum dependence as that of the momentum space {\it Zitterbewegung}.

Besides the above physical consequences, the dynamics governed by Majorana equation also shows unphysical phenomena. For example, the fidelity $\left|\left\langle\psi\left(t\right)|\psi_\theta\left(t\right)\right\rangle\right|^2$, where $\left|\psi\left(t\right)\right\rangle$ and $\left|\psi_\theta\left(t\right)\right\rangle$ are two Majorana spinors that evolve from initial states differing only in a global phase, $\left|\psi_\theta\left(0\right)\right\rangle=e^{i\theta}\left|\psi\left(0\right)\right\rangle$, will not always be unity as shown in Fig~\ref{fig:Fig2} ({\bf C}). In other words, a Majorana spinor does not have the freedom to choose an arbitrary global phase. The reason for this surprising effect is the existence of the complex conjugation $\hat K$ in the Majorana equation in Eq. (\ref{eq:MajoranaEquationOriginal}). This effect can be more explicitly shown in the mapping ${\mathcal M}$ in Eq. (\ref{eq:Mapping}), i.e. the global phase actually changes the initial four-component spinor of Eq. (\ref{eq:MajoranaEquationEnlarged}) in the enlarged Hilbert space. Figs.~\ref{fig:Fig2} ({\bf E}) and ({\bf F}) show an example of the experimental results of the density matrices in the enlarged and original Hilbert spaces, which are indeed different from each other. In Fig.~\ref{fig:Fig2} ({\bf D}), we experimentally observe the non-conservation of the orthogonality defined as $\left|\left\langle\psi\left(t\right)|\psi_\perp\left(t\right)\right\rangle\right|^2$, with $\left|\psi_\perp\left(t\right)\right\rangle$  being the Majorana spinor evolved from an orthogonal initial state $\left(\begin{smallmatrix} 0 \\ 1 \end{smallmatrix}\right)\otimes\left|-p\right\rangle$. During the evolution, the initial Majorana spinor will be coupled to $\left(\begin{smallmatrix} 0 \\ 1 \end{smallmatrix}\right)\otimes\left|p\right\rangle$ through the Hermitian relativistic kinetic term $\hat\sigma_x\hat p_x$, and $\left(\begin{smallmatrix} 0 \\ 1 \end{smallmatrix}\right)\otimes\left|-p\right\rangle$ through the non-Hermitian Majorana mass term $-im\hat\sigma_y\hat K$. The orthogonality $\left\langle\psi\left(t\right)|\psi'_\perp\left(t\right)\right\rangle$, where $\left|\psi'_\perp\left(t\right)\right\rangle$ is the Majorana spinor that evolves from the initial state $\left(\begin{smallmatrix} 0 \\ 1 \end{smallmatrix}\right)\otimes\left|p\right\rangle$, is always zero. This clearly indicates that the non-conservation of the orthogonality $\left|\left\langle\psi\left(t\right)|\psi_\perp\left(t\right)\right\rangle\right|^2$ stems from the non-Hermitian part of the Majorana Hamiltonian. As a result, given the same Majorana mass, we understand that the amplitude of the orthogonality oscillation is inversely proportional to the initial momentum.

\begin{figure}[htbp]
  \includegraphics[width=\textwidth]{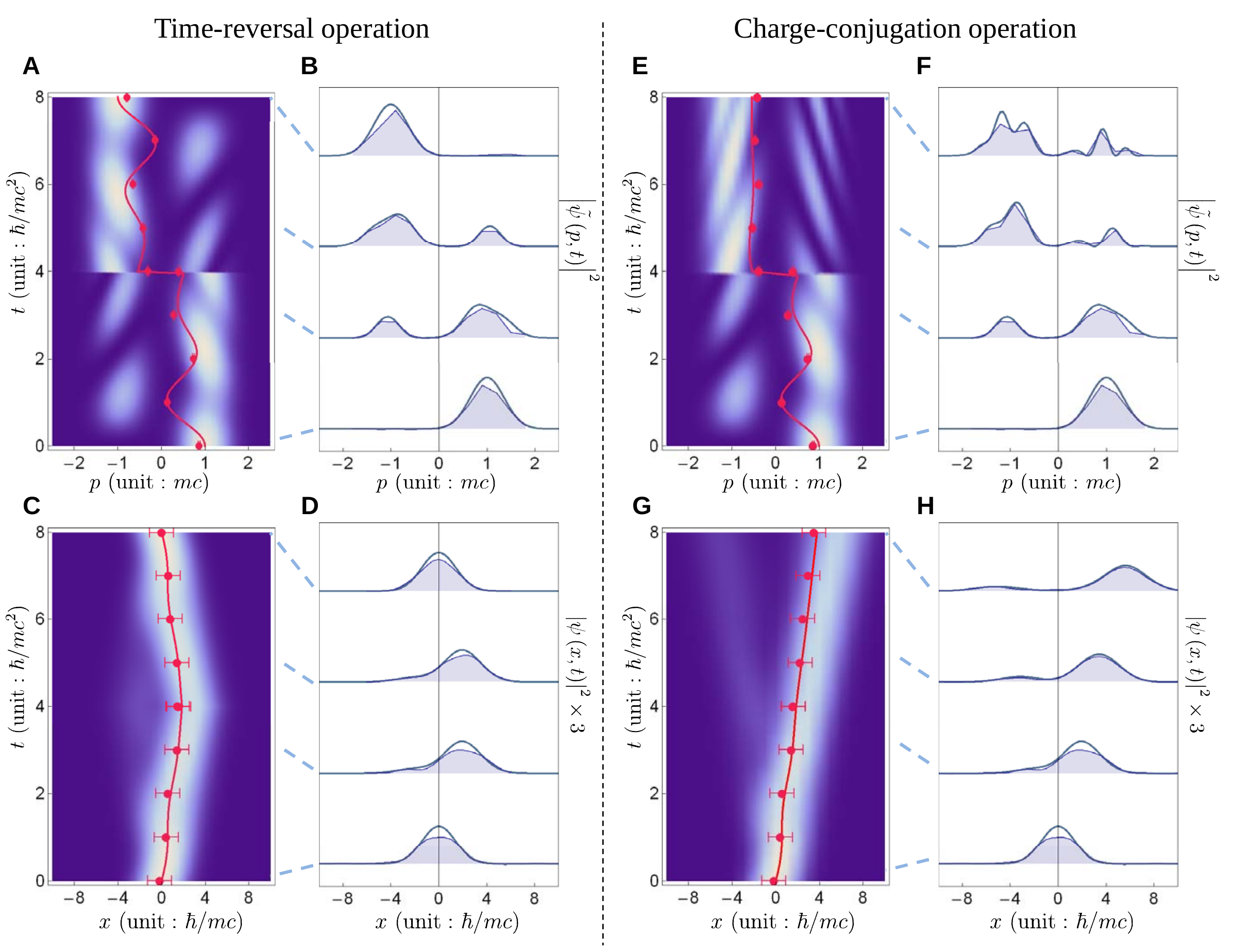}\\
  \caption[Text excluding the matrix]{\textbf{Time reversal and charge conjugation.} ({\bf A}-{\bf D}) Time reversal during the Majorana evolution. ({\bf E}-{\bf H}) Charge conjugation on top of Majorana dynamics. The initial state is a moving Gaussian wave packet $\psi\left(x,t=0\right)=\frac{1}{2}\pi^{-1/4}e^{-x^2/8-ip_0x}\left(\protect\begin{smallmatrix}1\\1\protect\end{smallmatrix}\right)$ with initial average momentum $p_0=1$. The evolution is governed by the Majorana equation and the symmetry operations are performed at the midpoint $t=4$. ({\bf A}, {\bf E}) The time-dependent density distributions in the momentum space. ({\bf C}, {\bf G}) The time-dependent density distributions in the position space. The solid curve represents the theoretical results of the average momentum $\left\langle\hat p_x\right\rangle$ in ({\bf A}, {\bf C}) and the average position $\left\langle\hat x\right\rangle$ in ({\bf E}, {\bf G}), while the red dots with error bars are experimental results. ({\bf B}, {\bf F}) The density distributions in the momentum space at given times. ({\bf D}, {\bf H}) The density distributions in the position space at given times. From bottom to top, the times are $t=0$, $3$, $5$ and $8$, respectively. The Solid curves are from theoretical calculation while the shades are from experiment. The curves are displaced along the vertical axis for better visualization, and the distributions in the position space are amplified by 3 times. The corresponding data points are marked by blue dashed lines.\label{fig:Fig3}}
\end{figure}

Other than the plane waves, we also implement Majorana dynamics with realistic initial wave packets in our embedding quantum simulator. For example, the initial states for the Majorana dynamics in Fig.~\ref{fig:Fig3} are moving Gaussian states with momentum distributions centered around $p_0=1$ with internal state $\frac{1}{\sqrt{2}}\left(\begin{smallmatrix} 1 \\ 1 \end{smallmatrix}\right)$. The first part of the time axis ($0\leq t<4$) in Fig.~\ref{fig:Fig3} represents the Majorana dynamics of a moving wave packet, where we observe damping oscillation in the momentum space and {\it Zitterbewegung} in the position space. The reason of the damping in the momentum space is that a Gaussian wave packet has distribution over many different momentum components, and each momentum component oscillates with different frequency. To our surprise, although the average momentum of a Majorana particle behaves quite different from that of a Dirac particle, there is no visible difference in the behaviors of the average position as well as the probability distribution in position space. This is because a Majorana particle oscillates between the particle and antiparticle components with inverse momentum, but the positions as well as the velocities of the particle and antiparticle are exactly the same \cite{SakuraiAdvanced}.

During the evolution of the Majorana equation, we implement the antiunitary time-reversal and charge-conjugation operations. Figs.~\ref{fig:Fig3} ({\bf A}-{\bf D}) show our experimental results of the time-reversal operation during the Majorana time evolution. As shown in Fig.~\ref{fig:Fig3} ({\bf A}), right after the time-reversal operation, the momentum as well as the velocity changes sign. As a result, the direction of the wave packet is reversed as shown in Fig. \ref{fig:Fig3} ({\bf C}). The damped average momentum as well as the position center of the wave packet is revived, which clearly shows that time is indeed reversed. Figs.~\ref{fig:Fig3} ({\bf E}-{\bf F}) demonstrate the experimental implementation of the charge-conjugation operation. The latter interchanges the particle and antiparticle components, which are defined from the corresponding Dirac equation with the same momentum as discussed in Fig. \ref{fig:Fig2} ({\bf B}). By definition, the particle and corresponding antiparticle have opposite momentum but the same velocity. As a result, right after the charge-conjugation operation, the average momentum is reversed but not the velocity. Therefore, the trajectory in position space remains intact, which is different from the time-reversal operation.

The demonstrated symmetry operations, as well as the $1+1$ Majorana dynamics, can be straightforwardly scaled up to several $3+1$ particles. The embedding quantum simulator for multipartite systems can be constructed by doubling the original Hilbert space dimension, which can be easily achieved by replacing only one two-level system of a coupled two-level system array by a four-level one. The proposed embedding scheme for the implementation of the time reversal and the charge conjugation operations may be extended for parity symmetry operation~\cite{Rodriguez13}. This enhanced toolbox for quantum simulators will be valuable for studying conservation laws and improving the computational capabilities of current quantum platforms.

\vspace*{0.5cm}

\section*{Acknowledgement}
This work was supported by the National Basic Research Program of China under Grants No. 2011CBA00300 (No. 2011CBA00301), the National Natural Science Foundation of China 61073174, 61033001, 61061130540, and 11374178, the Basque Government IT472-10 Grant, Spanish MINECO FIS2012-36673-C03-02, Ram\'on y Cajal Grant RYC-2012-11391, UPV/EHU UFI 11/55, CCQED, PROMISCE, SCALEQIT European projects, and the Alexander von Humboldt research grant. M.-H.Y. and K.K. acknowledge the recruitment program of global youth experts of China.

\clearpage

\section*{Supplementary Materials}

\setcounter{equation}{0}
\setcounter{figure}{0}
\setcounter{table}{0}
\setcounter{page}{1}
\makeatletter
\renewcommand{\theequation}{S\arabic{equation}}
\renewcommand{\thefigure}{S\arabic{figure}}

\section{Mapping between original and enlarged spaces}

In order to implement unphysical operations, such as the time reversal and charge conjugation, in our prototype of the embedding quantum simulator, we first consider the mapping ${\mathcal M}$ which transforms the state vector in the original $n$-dimensional complex Hilbert space ${\mathbb C}^n$ onto an enlarged $2n$-dimensional real Hilbert space ${\mathbb R}^{2n}$,
\begin{eqnarray}
\psi\left(x\right)=\left(\begin{array}{c} \psi_1\left(x\right) \\ \vdots \\ \psi_n\left(x\right) \end{array}\right)\xlongrightarrow{\mathcal M}\Psi\left(x\right)=\frac{1}{2}\left(\begin{array}{c} \psi_1\left(x\right)+\psi_1^*\left(x\right) \\ \vdots \\ \psi_n\left(x\right)+\psi_n^*\left(x\right) \\ i\left(\psi_1^*\left(x\right)-\psi_1\left(x\right)\right) \\ \vdots \\ i\left(\psi_n^*\left(x\right)-\psi_n\left(x\right)\right) \end{array}\right)\equiv\left(\begin{array}{c} \psi_1^{\rm re}\left(x\right) \\ \vdots \\ \psi_n^{\rm re}\left(x\right) \\ \psi_1^{\rm im}\left(x\right) \\ \vdots \\ \psi_n^{\rm im}\left(x\right) \end{array}\right).
\end{eqnarray}

In the $1+1$ dimension case, we consider the specific mapping ${\mathcal M}:{\mathbb C}^2\rightarrow{\mathbb R}^4$. In the following, we use a plane-wave initial state $\psi_p\left(x\right)$ as an example of the encoding of states in the enlarged Hilbert space,
\begin{eqnarray}
\psi_p\left(x\right)&=&\left(\begin{array}{c} C_1 \\ C_2 \end{array}\right)\otimes\frac{1}{\sqrt{2\pi}}e^{ipx/\hbar}\\
\xlongrightarrow{\mathcal M}\Psi_p\left(x\right)&=&\frac{1}{2}\left(\begin{array}{c} C_1^* \\ C_2^* \\ iC_1^* \\ iC_2^* \end{array}\right)\otimes\frac{1}{\sqrt{2\pi}}e^{-ipx/\hbar}+\frac{1}{2}\left(\begin{array}{c} C_1 \\ C_2 \\ -iC_1 \\ -iC_2 \end{array}\right)\otimes\frac{1}{\sqrt{2\pi}}e^{ipx/\hbar}\nonumber\\
&\equiv&\Psi_p^{(-)}\left(x\right)+\Psi_p^{(+)}\left(x\right),\nonumber
\end{eqnarray}
where $\Psi_p^{\pm}\left(x\right)$ corresponds to plane-wave states (unnormalized) with momentum $\pm p$. Here we want to emphasize two points: (i) although $\Psi_p\left(x\right)$ is real, the components $\Psi_p^{\pm}\left(x\right)$ are usually composed of complex functions; (ii) there are always $+p$ and $-p$ components in the enlarged space to guarantee $\Psi_p\left(x\right)$ is real.

\section{Quasi-quantum treatment of the momentum}

The $1+1$ Majorana equation for a two-component complex spinor $\psi\left(x\right)=\left(\begin{smallmatrix} \psi_1\left(x\right) \\ \psi_2\left(x\right) \end{smallmatrix}\right)$ is mapped onto a $3+1$ Dirac equation for a four-component real bispinor $\Psi\left(x\right)=\left(\begin{smallmatrix} \psi_1^{\rm re}\left(x\right) \\ \psi_2^{\rm re}\left(x\right) \\ \psi_1^{\rm im}\left(x\right) \\ \psi_2^{\rm im}\left(x\right) \end{smallmatrix}\right)$ in the enlarged space, which takes the following form,
\begin{eqnarray}
i\hbar\partial_t\Psi\left(x, t\right)=\hat{\mathcal H}\Psi\left(x, t\right)=\left[c\hat p_x\left(\hat{\mathbb I}\otimes\hat\sigma_x\right)-mc^2\left(\hat\sigma_x\otimes\hat\sigma_y\right)\right]\Psi\left(x, t\right),
\end{eqnarray}
where $\hat p_x=-i\hbar\partial_x$ is the momentum operator.

In the momentum space, the bispinor $\widetilde\Psi\left(p,t\right)$ is obtained via the Fourier transformation,
\begin{eqnarray}
\widetilde\Psi\left(p,t\right)=\frac{1}{\sqrt{2\pi}}\int\Psi\left(x,t\right)e^{-ipx/\hbar}dx,
\end{eqnarray}
and the equation of motion becomes
\begin{eqnarray}
i\hbar\partial_t\widetilde\Psi\left(p,t\right)=\hat{\mathcal H}_p\widetilde\Psi\left(p,t\right)=\left[cp\left(\hat{\mathbb I}\otimes\hat\sigma_x\right)-mc^2\left(\hat\sigma_x\otimes\hat\sigma_y\right)\right]\widetilde\Psi\left(p, t\right),
\end{eqnarray}
where the momentum operator $\hat p_x$ is substituted by its eigenvalue $p$. It is clear that the dynamics governed by $\hat{\mathcal H}_p$ is ready to be implemented in a quantum four-level system.

\section{Microwave Raman transitions}
In our $\Yb$ ion system, we use the microwaves for the transitions between $\mbox{\ensuremath{\ket 1}}$ and $\ket 2,\ket 3,\ket 4$ levels. We use a microwave Raman scheme similar to the widely used Raman laser scheme for the transitions between $\ket 2$ to $\ket 3$ and $\ket 3$ to $\ket 4$ transitions. We cannot apply a radio frequency for the operations of these transitions, since the energy gap between $\ket 2\leftrightarrow\ket 3$ and $\ket 3\leftrightarrow\ket 4$ is very close, which is $(2\pi)31$ kHz for our experimental condition. As shown in Fig. 1 of the main text, 6 different frequencies of microwaves are combined and simultaneously applied to the trap. For the control of 6 microwaves, we use a PCI-board arbitrary waveform generator (AWG) with $1$ GHz sampling rate, which is mixed with a $12442.8213$ MHz microwave. The AWG generates the signal of 6 frequencies from $186\sim214$ MHz.

The system is described by the Hamiltonian $\hat H=\hat H_A+\hat H_{AL}$, with atomic part $\hat H_A$ being
\begin{eqnarray}
\hat H_A=\left(\hbar\omega_{\rm hf}-\hbar\omega_z\right)\left|2\right\rangle\left\langle 2\right|+\left(\hbar\omega_{\rm hf}+\omega_q\right)\left|3\right\rangle\left\langle 3\right|+\left(\hbar\omega_{\rm hf}+\hbar\omega_z\right)\left|4\right\rangle\left\langle 4\right|,
\end{eqnarray}
and the interaction part $\hat H_{AL}$ being
\begin{eqnarray}
\hat H_{AL}\left(t\right)=\sum_{n=1}^6\sum_{j=2}^4\hbar\Omega_{1j}^{(n)}\cos\left(\omega_nt+\phi_n\right)\hat\sigma_{x}^{(j)},
\end{eqnarray}
respectively. We set the 6 frequencies in the microwave as follows,
\begin{eqnarray}
\begin{array}{lll}
\omega_1=\omega_{\rm hf}-\omega_z-\delta_1, &\quad& \omega_2=\omega_{\rm hf}+\omega_z-\delta_2,\\
\omega_3=\omega_{\rm hf}-\omega_z-\Delta, &\quad& \omega_4=\omega_{\rm hf}+\omega_q-\Delta-\delta_3,\\
\omega_5=\omega_{\rm hf}+\omega_q+\Delta, &\quad& \omega_6=\omega_{\rm hf}+\omega_z+\Delta-\delta_4,
\end{array}
\end{eqnarray}
where $\Delta$ is the detuning for the stimulated Raman transitions and $\delta_{i}$ are the frequency shifts used to compensate the AC Stark effect.
Using the method in Ref. \cite{James07}, we obtain the effective Hamiltonian $\hat{\mathcal H}_{\rm eff}=\hat{\mathcal H}_{\rm st}+\hat{\mathcal H}_{\rm cp}$ in the interaction picture defined by $\hat H_A$, where $\hat{\mathcal H}_{\rm st}$ includes all of the AC Stark shifts,
\begin{eqnarray}
\hat{\mathcal H}_{\rm st}&=&\frac{\hbar\left(\Omega_{13}^{(1)}\right)^2}{4\left(\omega_z+\omega_q+\delta_1\right)}\left(|3\rangle\langle3|-|1\rangle\langle1|\right) +\frac{\hbar\left(\Omega_{14}^{(1)}\right)^2}{4\left(2\omega_z+\delta_1\right)}\left(|4\rangle\langle4|-|1\rangle\langle1|\right)\nonumber\\
&&-\frac{\hbar\left(\Omega_{12}^{(2)}\right)^2}{4\left(2\omega_z-\delta_2\right)}\left(|2\rangle\langle2|-|1\rangle\langle1|\right) -\frac{\hbar\left(\Omega_{13}^{(2)}\right)^2}{4\left(\omega_z-\omega_q-\delta_2\right)}\left(|3\rangle\langle3|-|1\rangle\langle1|\right)\nonumber\\
&&+\frac{\hbar\left(\Omega_{12}^{(3)}\right)^2}{4\Delta}\left(|2\rangle\langle2|-|1\rangle\langle1|\right) +\frac{\hbar\left(\Omega_{13}^{(3)}\right)^2}{4\left(\omega_z+\Delta+\omega_q\right)}\left(|3\rangle\langle3|-|1\rangle\langle1|\right) \nonumber  \\ && + \frac{\hbar\left(\Omega_{14}^{(3)}\right)^2}{4\left(2\omega_z+\Delta\right)}\left(|4\rangle\langle4|-|1\rangle\langle1|\right)\nonumber\\
&&-\frac{\hbar\left(\Omega_{12}^{(4)}\right)^2}{4\left(\omega_z-\Delta+\omega_q-\delta_3\right)}\left(|2\rangle\langle2|-|1\rangle\langle1|\right) +\frac{\hbar\left(\Omega_{13}^{(4)}\right)^2}{4\left(\Delta+\delta_3\right)}\left(|3\rangle\langle3|-|1\rangle\langle1|\right) \nonumber\\ &&+\frac{\hbar\left(\Omega_{14}^{(4)}\right)^2}{4\left(\omega_z+\Delta-\omega_q+\delta_3\right)}\left(|4\rangle\langle4|-|1\rangle\langle1|\right)\nonumber\\
&&-\frac{\hbar\left(\Omega_{12}^{(5)}\right)^2}{4\left(\omega_z+\Delta+\omega_q\right)}\left(|2\rangle\langle2|-|1\rangle\langle1|\right) -\frac{\hbar\left(\Omega_{13}^{(5)}\right)^2}{4\Delta}\left(|3\rangle\langle3|-|1\rangle\langle1|\right) \nonumber\\ &&+\frac{\hbar\left(\Omega_{14}^{(5)}\right)^2}{4\left(\omega_z-\Delta-\omega_q\right)}\left(|4\rangle\langle4|-|1\rangle\langle1|\right)\nonumber\\
&&-\frac{\hbar\left(\Omega_{12}^{(6)}\right)^2}{4\left(\omega_z+\Delta-\delta_4\right)}\left(|2\rangle\langle2|-|1\rangle\langle1|\right) -\frac{\hbar\left(\Omega_{13}^{(6)}\right)^2}{4\left(\omega_z+\Delta-\omega_q-\delta_4\right)}\left(|3\rangle\langle3|-|1\rangle\langle1|\right) \nonumber\\ &&-\frac{\hbar\left(\Omega_{14}^{(6)}\right)^2}{4\left(\Delta-\delta_4\right)}\left(|4\rangle\langle4|-|1\rangle\langle1|\right)\nonumber\\
&=&\hbar\omega_{\rm st}^{(1)}|1\rangle\langle1|+\hbar\omega_{\rm st}^{(2)}|2\rangle\langle2|+\hbar\omega_{\rm st}^{(3)}|3\rangle\langle3|+\hbar\omega_{\rm st}^{(4)}|4\rangle\langle4|,
\end{eqnarray}
and $\hat{\mathcal H}_{\rm cp}$ includes all coupling terms with frequencies much smaller than $\Delta$,
\begin{eqnarray}
\hat{\mathcal H}_{\rm cp}\!\!\!\!&=&\!\!\!\!\frac{\hbar\widetilde\Omega_{12}e^{-i\delta_1t+i\phi_1}}{2}|1\rangle\langle2| +\frac{\hbar\widetilde\Omega_{14}e^{-i\delta_2t+i\phi_2}}{2}|1\rangle\langle4|\\
&&\!\!\!+\frac{\hbar\widetilde\Omega_{23}e^{-i\delta_3t+i\phi_{43}}}{2}|2\rangle\langle3|\!+\!\frac{\hbar\widetilde\Omega_{34}e^{-i\delta_4t+i\phi_{65}}}{2}|3\rangle\langle4|\nonumber\\
&&\!\!\!\!+\frac{\hbar\Omega_{13}^{(3)}\Omega_{14}^{(4)}\left(\omega_z+\Delta\right)}{4\left[\left(\omega_z+\Delta\right)^2-\omega_q^2\right]} |3\rangle\langle4|e^{i\left(2\omega_q-\delta_3\right)t+i\phi_{43}}
-\frac{\hbar\Omega_{12}^{(5)}\Omega_{13}^{(6)}\left(\omega_z+\Delta\right)}{4\left[\left(\omega_z+\Delta\right)^2-\omega_q^2\right]} |2\rangle\langle3|e^{-i\left(2\omega_q+\delta_4\right)t+i\phi_{65}}+{\rm H.c.}\nonumber
\end{eqnarray}
The effective couplings $\widetilde\Omega_{23}$ and $\widetilde\Omega_{34}$ are defined as follows,
\begin{eqnarray}
\widetilde\Omega_{23}=\frac{\Omega_{12}^{(3)}\Omega_{13}^{(4)}}{4}\left(\frac{1}{\Delta}+\frac{1}{\Delta+\delta_3}\right),\quad \widetilde\Omega_{34}=-\frac{\Omega_{13}^{(5)}\Omega_{14}^{(6)}}{4}\left(\frac{1}{\Delta}+\frac{1}{\Delta-\delta_4}\right).
\end{eqnarray}
Transferring into the second rotating frame defined by $\hat{\mathcal H}_{\rm st}$, one obtains the following rotating frame Hamiltonian
\begin{eqnarray}
\hat{\mathcal H}=e^{i\hat{\mathcal H}_{\rm st}t/\hbar}\hat{\mathcal H}_{\rm cp}\left(t\right)e^{-i\hbar{\mathcal H}_{\rm st}t/\hbar}.
\end{eqnarray}
In order to make the final rotating frame Hamiltonian time-independent, the additional detunings should satisfy the following relations,
\begin{eqnarray}
\delta_1&=&\omega_{\rm st}^{(1)}-\omega_{\rm st}^{(2)},\\
\delta_2&=&\omega_{\rm st}^{(1)}-\omega_{\rm st}^{(4)},\nonumber\\
\delta_3&=&\omega_{\rm st}^{(2)}-\omega_{\rm st}^{(3)},\nonumber\\
\delta_4&=&\omega_{\rm st}^{(3)}-\omega_{\rm st}^{(4)}.\nonumber
\end{eqnarray}
Comparing to the Majorana equation, one obtains the following relations
\begin{eqnarray}
&&\frac{\hbar\widetilde\Omega_{12}}{2}e^{i\phi_1}=\frac{\hbar\widetilde\Omega_{34}}{2}e^{i\phi_{65}}=cp,\\
&&\frac{\hbar\widetilde\Omega_{14}}{2}e^{i\phi_2}=-\frac{\hbar\widetilde\Omega_{23}}{2}e^{i\phi_{43}}=imc^2.\nonumber
\end{eqnarray}

One problem in this scheme is the slowing down of operations. The Raman transition is $10$ times slower than normal Rabi flopping, even with full power. But now we need 6 microwaves together. Decoherence occurs when the whole microwave duration is longer than $600\mu$s. This decoherence problem is later solved by applying a line trigger \cite{Smith11} to the pulse sequencer.

\section{Physical observables}

In this part, we will describe in detail the procedure to extract the information of various physical observables from experimental data. The time-dependent enlarged four-component spinor can be formally written in the momentum space as follows,
\begin{eqnarray}
\left|\Psi\left(t\right)\right\rangle=\int\Psi\left(p\right)\left|\chi_p\left(t\right)\right\rangle\otimes\left|p\right\rangle,\label{eq:EnlargedSpinor}
\end{eqnarray}
with $\left|p\right\rangle$ being the momentum basis, i.e. the plane-wave states, and $\left|\chi_p\left(t\right)\right\rangle$ describing the internal state $\left|\chi_p\left(t\right)\right\rangle=\sum_{j=1}^4\chi_{p,j}\left(t\right)\left|j\right\rangle$. Note that the wave function in the momentum space $\Psi\left(p\right)$ does not depend on time and is fully determined by the initial condition $\left|\psi\left(0\right)\right\rangle$ in the original space, because the effective Hamiltonian $\hat{\mathcal H}$ commutes with the momentum operator $\hat p_x$. The only time-dependent part in Eq. (\ref{eq:EnlargedSpinor}) is the internal state $\left|\chi_p\left(t\right)\right\rangle$, whose dynamics is determined by the enlarged space Hamiltonian $\hat{\mathcal H}_p=pc\left(\ket{1}\bra{2}+\ket{3}\bra{4}\right)+i mc^2 \left(\ket{1}\bra{4}-\ket{2}\bra{3}\right) + {\rm H.c.}$. The equation of motion for $\left|\chi_p\left(t\right)\right\rangle$, given by $i\hbar\partial_t\left|\chi_p\left(t\right)\right\rangle=\hat{\mathcal H}_p\left|\chi_p\left(t\right)\right\rangle$, can be simulated in a quantum four-level system. Using quantum state tomography, we experimentally obtain the density matrix $\hat\varrho_p\left(t\right)$ corresponding to $\left|\chi_p\left(t\right)\right\rangle\left\langle\chi_p\left(t\right)\right|$.

\subsection{Diagonal operators in the momentum space}

The general form of a diagonal operator $\hat{\mathcal O}_{\rm dg}$ in the momentum space can be written as follows,
\begin{eqnarray}
\hat{\mathcal O}_{\rm dg}=\hat\Sigma\otimes f\left(\hat p\right),
\end{eqnarray}
with $\hat\Sigma=c_0\hat{\mathbb I}+c_1\hat\sigma_x+c_2\hat\sigma_y+c_3\hat\sigma_z$ and $f\left(\cdot\right)$ being an arbitrary algebraic function, $\left\langle p\left|f\left(\hat p\right)\right|p'\right\rangle=f\left(p\right)\delta\left(p-p'\right)$.
The expectation value of this operator at arbitrary time $t$ can be obtained as follows,
\begin{eqnarray}
\left\langle\Psi\left(t\right)\left|M^\dag\hat{\mathcal O}_{\rm dg}M\right|\Psi\left(t\right)\right\rangle&=&\int dp dp'\Psi^*\left(p\right)\Psi\left(p'\right)\left\langle p\left|f\left(\hat p\right)\right|p'\right\rangle\left\langle\chi_p\left(t\right)\left|M^\dag\hat\Sigma M\right|\chi_{p'}\left(t\right)\right\rangle\nonumber\\
&=&\int dp\left|\Psi\left(p\right)\right|^2f\left(p\right){\rm Tr}\left[\hat\varrho_p\left(t\right)\hat M^\dag\hat\Sigma\hat M\right].
\end{eqnarray}
We may take the average momentum as a simple example,
\begin{eqnarray}
p\left(t\right)&\equiv&\left\langle\psi\left(t\right)\left|\hat p\right|\psi\left(t\right)\right\rangle\\
&=&\left\langle\Psi\left(t\right)\left|M^\dag\hat pM\right|\Psi\left(t\right)\right\rangle\nonumber\\
&=&\int dpp\left|\Psi\left(p\right)\right|^2{\rm Tr}\left[\hat\varrho_p\left(t\right)\hat M^\dag\hat M\right].\nonumber
\end{eqnarray}

The quantum simulation for each $\left|\chi_p\left(t\right)\right\rangle$ will be as follows.
\begin{enumerate}
\item Prepare the initial state $\left|\chi_p\left(0\right)\right\rangle=\sum_{j=1}^4\chi_{p,j}\left(t\right)\left|j\right\rangle$.
\item Implement the Hamiltonian $\hat{\mathcal H}_p$ and let the system evolve for certain time duration $t$,
\begin{eqnarray}
\hat{\mathcal H}_p=\left(\begin{array}{cccc} 0 & cp & 0 & imc^2 \\ cp & 0 & -imc^2 & 0 \\ 0 & imc^2 & 0 & cp \\ -imc^2 & 0 & cp & 0 \end{array}\right).
\end{eqnarray}
\item Perform the quantum state tomography and obtain $\hat\varrho_p\left(t\right)=\left|\chi_p\left(t\right)\right\rangle\left\langle\chi_p\left(t\right)\right|$.
\end{enumerate}
Then the matrix element mentioned above can be obtained straightforwardly,
\begin{eqnarray}
\left\langle\chi_p\left(t\right)\left|M^\dag\hat\Sigma M\right|\chi_p\left(t\right)\right\rangle={\rm Tr}\left[\hat\rho_p\left(t\right)M^\dag\hat\Sigma M\right].
\end{eqnarray}

\subsection{Off-diagonal operators in the momentum space}

Then we turn to investigate the method to obtain the expectation value of some off-diagonal operators $\hat{\mathcal O}_{\rm od}$ in the momentum space. We will take position-dependent operators as examples, i.e.,
\begin{eqnarray}
\hat{\mathcal O}_{\rm od}=\hat\Sigma\otimes f\left(\hat x\right).
\end{eqnarray}
As mentioned above, the expectation value can be written as
\begin{eqnarray}
&&\left\langle\Psi\left(t\right)\left|M^\dag\hat{\mathcal O}_{\rm dg}M\right|\Psi\left(t\right)\right\rangle\\
&=&\int dp dp'\Psi^*\left(p\right)\Psi\left(p'\right)\left\langle p\left|f\left(\hat x\right)\right|p'\right\rangle\left\langle\chi_p\left(t\right)\left|M^\dag\hat\Sigma M\right|\chi_{p'}\left(t\right)\right\rangle.\nonumber
\end{eqnarray}
Since $f\left(\hat x\right)$ is not diagonal in the momentum space, the above expression will involve off-diagonal matrix-element as $\left\langle\chi_p\left(t\right)\left|M^\dag\hat\Sigma M\right|\chi_{p'}\left(t\right)\right\rangle$. If we stick to the previous scheme, we will obtain two independent density matrices $\hat\varrho_p\left(t\right)$ and $\hat\varrho_{p'}\left(t\right)$, from which we can not construct the off-diagonal matrix element between two distinct momenta.

Inspired by the effective Hamiltonian $\hat{\mathcal H}_p$ for some definite momentum $p$,
\begin{eqnarray}
\hat{\mathcal H}_p&=&pc\left(\ket{1}\bra{2}+\ket{3}\bra{4}\right)+i mc^2 \left(\ket{1}\bra{4}-\ket{2}\bra{3}\right) + {\rm H.c.}\nonumber\\
&\equiv& pc\left(\hat{\mathbb I}\otimes\hat\sigma_x\right)-mc^2\left(\hat\sigma_x\otimes\hat\sigma_z\right),
\end{eqnarray}
we notice that the first qubit can be diagonalized in the $\hat\sigma_x$-basis. The quantum states and operators in the new basis $\left\{|+\rangle|0\rangle,|+\rangle|1\rangle,|-\rangle|0\rangle,|-\rangle|1\rangle\right\}$ and the old basis $\left\{|0\rangle|0\rangle,|0\rangle|1\rangle,|1\rangle|0\rangle,|1\rangle|1\rangle\right\}$, where $\left|0\right\rangle$ and $\left|1\right\rangle$ are the eigenstates of $\hat\sigma_z$, are related by the following transform matrix $\hat S$,
\begin{eqnarray}
\hat S=\frac{\sqrt{2}}{2}\left(\begin{array}{rrrr} 1 & 0 & 1 & 0 \\ 0 & 1 & 0 & 1 \\ 1 & 0 & -1 & 0 \\ 0 & 1 & 0 & -1 \end{array}\right),
\end{eqnarray}
where $\left|\pm\right\rangle\equiv\frac{1}{\sqrt{2}}\left(\left|0\right\rangle\pm\left|1\right\rangle\right)$ are eigenstates of $\hat\sigma_x$.
In other words, the equation of motion for $\left|\chi_p\left(t\right)\right\rangle$ can be written in the new basis as follows,
\begin{eqnarray}
i\hbar\frac{\partial}{\partial t}\left(\begin{array}{c} \chi_p^+\left(t\right) \\ \chi_p^-\left(t\right) \end{array}\right)=\left(\begin{array}{cc} \hat H_p^+ & 0 \\ 0 & \hat H_p^- \end{array}\right)\left(\begin{array}{c} \chi_p^+\left(t\right) \\ \chi_p^-\left(t\right) \end{array}\right) \label{eq:EqnMotionInNewB}
\end{eqnarray}
with
\begin{eqnarray}
\left(\begin{array}{cc} \hat H_p^+ & 0 \\ 0 & \hat H_p^- \end{array}\right)=\hat S^\dag\hat H\left(p\right)\hat S=\left(\begin{array}{cccc} 0 & pc+imc^2 & 0 & 0 \\ pc-imc^2 & 0 & 0 & 0 \\ 0 & 0 & 0 & pc-imc^2 \\ 0 & 0 & pc+imc^2 & 0 \end{array}\right)
\end{eqnarray}
and
\begin{eqnarray}
\left(\begin{array}{c} \chi_p^+\left(t\right) \\ \chi_p^-\left(t\right) \end{array}\right)=\left(\begin{array}{c} \chi_{p,1}^+\left(t\right) \\ \chi_{p,2}^+\left(t\right) \\ \chi_{p,1}^-\left(t\right) \\ \chi_{p,2}^-\left(t\right) \end{array}\right) =\hat S^\dag\left(\begin{array}{c} \chi_{p,1}\left(t\right) \\ \chi_{p,2}\left(t\right) \\ \chi_{p,3}\left(t\right) \\ \chi_{p,4}\left(t\right) \end{array}\right),
\end{eqnarray}
where $\chi_p^\pm\left(t\right)$ are column vectors with two entries and $\hat H_p^\pm$ are $2\times2$ matrices in the new basis. As shown in Eq.~(\ref{eq:EqnMotionInNewB}), we note that the dynamics for $\chi_p^\pm\left(t\right)$ are totally decoupled from each other, and can be separately simulated in quantum two-level systems. In order to obtain off-diagonal matrix elements between two distinct momenta $p$ and $p'$, we have to simulate $\chi_p\left(t\right)$ and $\chi_{p'}\left(t\right)$ coherently. We obtain the following equations of motion by rearranging Eq.~(\ref{eq:EqnMotionInNewB}),
\begin{eqnarray}
i\hbar\frac{\partial}{\partial t}\left(\begin{array}{c} \chi_p^+\left(t\right) \\ \chi_{p'}^+\left(t\right) \end{array}\right)=\left(\begin{array}{cc} \hat H_p^+ & 0 \\ 0 & \hat H_{p'}^+ \end{array}\right)\left(\begin{array}{c} \chi_p^+\left(t\right) \\ \chi_{p'}^+\left(t\right) \end{array}\right),\\
i\hbar\frac{\partial}{\partial t}\left(\begin{array}{c} \chi_p^-\left(t\right) \\ \chi_{p'}^-\left(t\right) \end{array}\right)=\left(\begin{array}{cc} \hat H_p^- & 0 \\ 0 & \hat H_{p'}^- \end{array}\right)\left(\begin{array}{c} \chi_p^-\left(t\right) \\ \chi_{p'}^-\left(t\right) \end{array}\right),\nonumber
\end{eqnarray}
which can be simulated in quantum four-level systems.

In the following investigation, we will use the average position $\left\langle\hat x\right\rangle\equiv\left\langle\psi\left(t\right)\left|\hat x\right|\psi\left(t\right)\right\rangle$ as an example. The detailed derivation is as follows,
\begin{eqnarray}
\left\langle\hat x\right\rangle&=&\left\langle\Psi\left(t\right)\left|M^\dag\hat xM\right|\Psi\left(t\right)\right\rangle\nonumber\\
&=&\int dpdp'\Psi^*\left(p\right)\Psi\left(p'\right)\left\langle p\left|\hat x\right|p'\right\rangle\left\langle\chi_p\left(t\right)\left|M^\dag M\right|\chi_{p'}\left(t\right)\right\rangle\nonumber\\
&=&\int dpdp'\Psi^*\left(p\right)\Psi\left(p'\right)\nonumber\\
&&\times\left\langle p\left|\hat x\right|p'\right\rangle\left[\left\langle\chi_p^+\left(t\right)|\chi_{p'}^+\left(t\right)\right\rangle +\left\langle\chi_p^-\left(t\right)|\chi_{p'}^-\left(t\right)\right\rangle-i\left(\left\langle\chi_p^+\left(t\right)|\chi_{p'}^-\left(t\right)\right\rangle -\left\langle\chi_p^-\left(t\right)|\chi_{p'}^+\left(t\right)\right\rangle\right)\right]\nonumber\\
&=&\int\frac{dxdpdp'}{2\pi\hbar}x\exp\left[-\left(p-p'\right)x/\hbar\right]\Psi^*\left(p\right)\Psi\left(p'\right)\left[\left\langle\chi_p^+\left(t\right)|\chi_{p'}^+\left(t\right)\right\rangle +\left\langle\chi_p^-\left(t\right)|\chi_{p'}^-\left(t\right)\right\rangle\right].\nonumber
\end{eqnarray}
The last line in the above equation is valid because of the following identity,
\begin{eqnarray}
\int\frac{dpdp'}{2\pi\hbar}\exp\left[-\left(p-p'\right)x/\hbar\right]\Psi^*\left(p\right)\Psi\left(p'\right)\left[\left\langle\chi_p^+\left(t\right)|\chi_{p'}^-\left(t\right)\right\rangle -\left\langle\chi_p^-\left(t\right)|\chi_{p'}^+\left(t\right)\right\rangle\right]=0,
\end{eqnarray}
which can be verified using $\rho_E\left(p\right)=\rho_E\left(-p\right)$ and $\chi_p^\pm\left(t\right)=\left[\chi_{-p}^\pm\left(t\right)\right]^*$.

The experiment procedure would be as follows.
\begin{enumerate}
\item Prepare the initial state determined by the initial condition $\left[\chi_{p,1}^\pm\left(0\right),\chi_{p,2}^\pm\left(0\right),\chi_{p',1}^\pm\left(0\right),\chi_{p',2}^\pm\left(0\right)\right]^T$.
\item Implement $\hat H^\pm_{p,p'}$ and let the system evolve for some time period $t$,
\begin{eqnarray}
\hat H^\pm_{p,p'}=\left(\begin{array}{cc} \hat H_p^\pm & 0 \\ 0 & \hat H_{p'}^\pm \end{array}\right)=\left(\begin{array}{cccc} 0 & pc\pm imc^2 & 0 & 0 \\ pc\mp imc^2 & 0 & 0 & 0 \\ 0 & 0 & 0 & p'\pm imc^2 \\ 0 & 0 & p'\mp imc^2 & 0 \end{array}\right).
\end{eqnarray}
\item Perform the quantum state tomography and obtain $\rho^\pm_{p,p'}$,
\begin{eqnarray}
\rho^\pm_{p,p'}=\left(\begin{array}{cccc} \left|\chi_{p,1}^\pm\right|^2 & \chi_{p,1}^\pm\left(\chi_{p,2}^+\right)^* & \chi_{p,1}^\pm\left(\chi_{p',1}^\pm\right)^* & \chi_{p,1}^\pm\left(\chi_{p',2}^\pm\right)^* \\ \chi_{p,2}^\pm\left(\chi_{p,1}^\pm\right)^* & \left|\chi_{p,2}^\pm\right|^2 & \chi_{p,2}^\pm\left(\chi_{p',1}^\pm\right)^* & \chi_{p,2}^\pm\left(\chi_{p',2}^\pm\right)^* \\ \chi_{p',1}^\pm\left(\chi_{p,1}^\pm\right)^* & \chi_{p',1}^\pm\left(\chi_{p,2}^\pm\right)^* & \left|\chi_{p',1}^\pm\right|^2 & \chi_{p',1}^\pm\left(\chi_{p',2}^\pm\right)^* \\ \chi_{p',2}^\pm\left(\chi_{p,1}^\pm\right)^* & \chi_{p',2}^\pm\left(\chi_{p,2}^\pm\right)^* & \chi_{p',2}^\pm\left(\chi_{p',1}^\pm\right)^* & \left|\chi_{p',2}^\pm\right|^2 \end{array}\right).
\end{eqnarray}
\end{enumerate}
Sweeping the momenta $p$ and $p'$ over all possible values, we would obtain all of the information that is needed to calculate the expectation value $x\left(t\right)\equiv\left\langle\psi\left(t\right)\left|\hat x\right|\psi\left(t\right)\right\rangle$. The number of separate simulations for different $\left(p,p'\right)$ pairs for both signs will be $N_P^2$, where $N_P$ is the number of points with which we discretize the momentum axis.

\section{Charge conservation and charge conjugation}

The non-Hermitian Majorana Hamiltonian does not have eigenstates. However, we can define the concepts of particle and antiparticle from the eigenstates of the corresponding Dirac Hamiltonian, which is obtained by substituting the Majorana mass term with the Dirac mass term. Under the same convention, the $1+1$ Dirac equation takes the following dimensionless form,
\begin{eqnarray}
i\partial_t\left|\psi\right\rangle=\left(\hat\sigma_x\hat p_x+m\hat\sigma_z\right)\left|\psi\right\rangle,
\end{eqnarray}
with the eigenvalues $\pm\sqrt{p^2+m^2}$ and the corresponding eigenstates
\begin{eqnarray}
\left|\psi_p^{(+)}\right\rangle&=&\frac{1}{\sqrt{2}\left(p^2+m^2\right)^{1/4}}\left(\begin{array}{c}\sqrt{\sqrt{p^2+m^2}+m} \\ \left(p/\left|p\right|\right)\sqrt{\sqrt{p^2+m^2}-m} \end{array}\right)\otimes\left|p\right\rangle,\\
\left|\psi_p^{(-)}\right\rangle&=&\frac{1}{\sqrt{2}\left(p^2+m^2\right)^{1/4}}\left(\begin{array}{c}\sqrt{\sqrt{p^2+m^2}-m} \\ -\left(p/\left|p\right|\right)\sqrt{\sqrt{p^2+m^2}+m} \end{array}\right)\otimes\left|p\right\rangle.\nonumber
\end{eqnarray}

Starting from an initial Majorana spinor $\left|\psi\left(0\right)\right\rangle=\left(\begin{smallmatrix} u_1\left(0\right) \\ u_2\left(0\right) \end{smallmatrix}\right)\otimes\left|p\right\rangle$, the time-dependent Majorana spinor can be formally written as follows,
\begin{eqnarray}
\left|\psi\left(t\right)\right\rangle=\left(\begin{array}{c} v_1\left(t\right) \\ v_2\left(t\right) \end{array}\right)\otimes\left|-p\right\rangle+\left(\begin{array}{c} u_1\left(t\right) \\ u_2\left(t\right) \end{array}\right)\otimes\left|p\right\rangle.\label{eq:ParticleAntiParticle}
\end{eqnarray}
Note that the appearance of the negative momentum component is originated from the charge conjugation in the Majorana mass term. By definition, the time-dependent charge is obtained as follows,
\begin{eqnarray}
C\left(t\right)=\left|\left\langle\psi_p^{(+)}|\psi\left(t\right)\right\rangle\right|^2+\left|\left\langle\psi_{-p}^{(+)}|\psi\left(t\right)\right\rangle\right|^2 -\left|\left\langle\psi_p^{(-)}|\psi\left(t\right)\right\rangle\right|^2-\left|\left\langle\psi_{-p}^{(-)}|\psi\left(t\right)\right\rangle\right|^2.
\end{eqnarray}
By setting $\left(\begin{smallmatrix} u_1\left(0\right) \\ u_2\left(0\right) \end{smallmatrix}\right)=\left(\begin{smallmatrix} 1 \\ 0 \end{smallmatrix}\right)$, we obtain the theoretical and experimental data shown in Fig. 2 (b) in the main text.

\begin{figure}[htbp]
  \includegraphics[width=\textwidth]{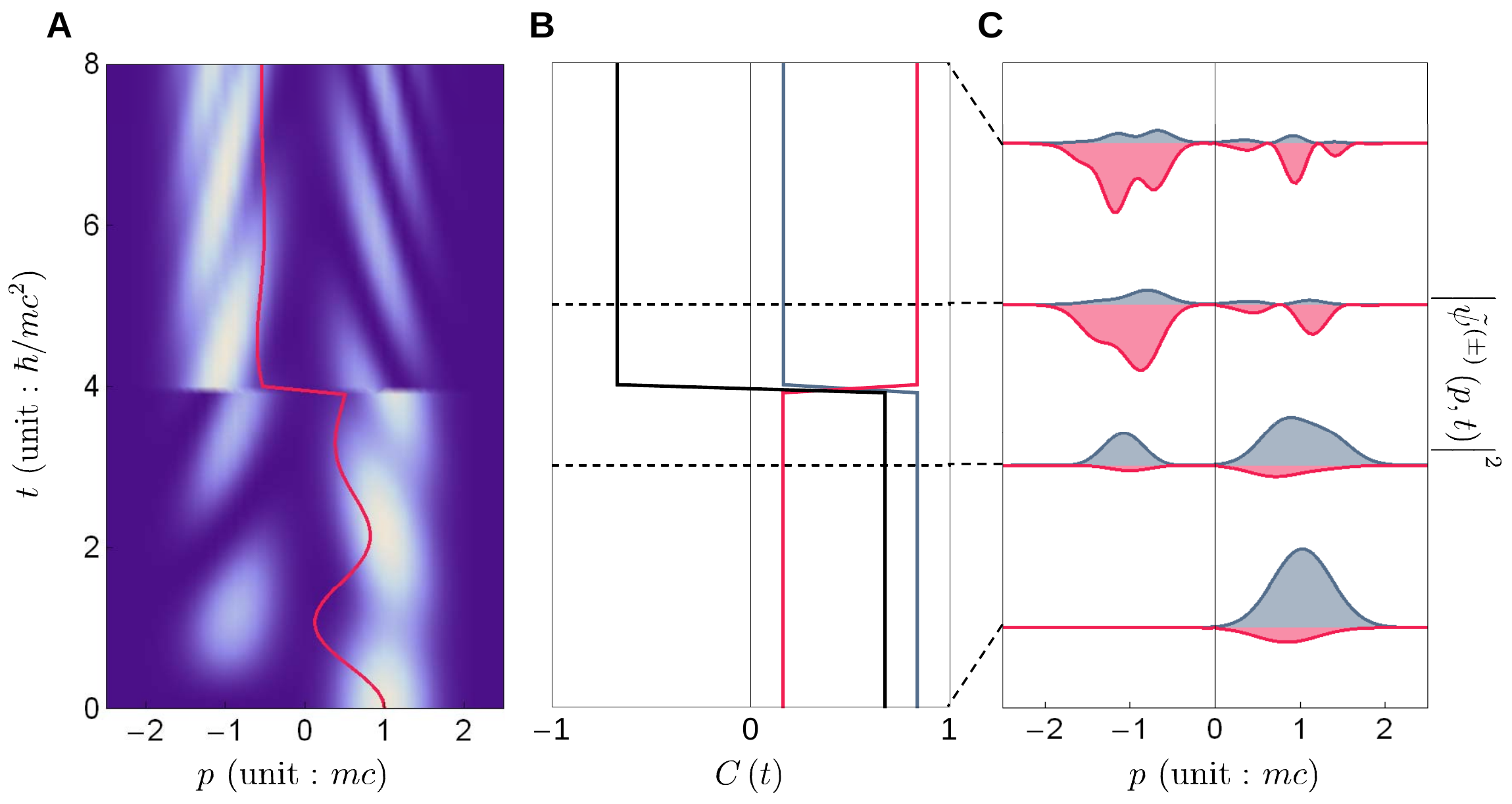}\\
  \caption[Text excluding the matrix]{{\bf Charge conjugation on top of the Majorana dynamics.} ({\bf A}) Time-dependent density distributions in momentum space. We implement the charge conjugation at the intermediate time $t=4$. The solid line represents the average value of the momentum. ({\bf B}) Time-dependent charge (black) as well as populations of the particle (blue) and antiparticle (red) components. ({\bf C}) Momentum distributions of the particle (blue) and antiparticle (red) components at different times $t=0$, $3$, $5$, and $8$. The curves are displaced along the vertical axis for better visualization.\label{fig:FigS1}}
\end{figure}

Besides the violation of the charge conservation of plane-wave initial states, we investigate the charge conjugation on top of the Majorana dynamics of an initial moving Gaussian wave packet as shown in Figs. 3 ({\bf E}-{\bf H}) in the main text. The initial Majorana spinor takes form of $\psi\left(x,t=0\right)=\frac{1}{2}\pi^{-1/4}e^{-x^2/8-ip_0x}\left(\begin{smallmatrix} 1 \\ 1 \end{smallmatrix}\right)$ with $p_0=1$ in the position space. By definition, the charge conjugation interchanges the particle and antiparticle components in Eq. (\ref{eq:ParticleAntiParticle}). In addition to the results in the main text, here we show the theoretical result for the dynamics of the internal degree of freedom in Fig. \ref{fig:FigS1}. In Fig. \ref{fig:FigS1} ({\bf B}), we can clearly see that the populations of the particle and antiparticle components are interchanged right after the implementation of the charge-conjugation operator. Fig. \ref{fig:FigS1} ({\bf C}) shows the momentum distributions for the particle and antiparticle components at different times above and below the base lines, respectively. We clearly see from the Majorana dynamics of the internal degree of freedom that the evolution is continued after the implementation of the charge conjugation, although the roles of the particle and antiparticle are interchanged.

\end{document}